\documentclass{an}
\usepackage{graphicx}
\usepackage{times}
\usepackage{fancyhdr}
\def\spose#1{\hbox to 0pt{#1\hss}}
\def\oo{[O\textsc{ii}]\,}
\def\s2{[S\textsc{ii}]}
\def\o3{[O\textsc{iii}]}
\def\ne3{[Ne\textsc{iii}]}
\def\lta{\mathrel{\spose{\lower 3pt\hbox{$\mathchar"218$}}\raise 2.0pt\hbox{$\ma
thchar"13C$}}}
\def\gta{\mathrel{\spose{\lower 3pt\hbox{$\mathchar"218$}}\raise 2.0pt\hbox{$\ma
thchar"13E$}}}
\def\arcsec{$^{\prime\prime}$}

\begin{document}

\title{Deep spectroscopy of a young radio source at z=0.521}

\author{K.J. Inskip\inst{1}, D. Lee\inst{2}, Garret Cotter\inst{2},
  A.C.S. Readhead\inst{3} \& T.J.Pearson\inst{3}}
\institute{Department of Physics \& Astronomy, University of
  Sheffield, Sheffield S3 7RH, UK
\and Astrophysics, Denys Wilkinson Building, Keble Road, Oxford OX1
  3RH, UK
\and California Institute of Technology, Pasadena, CA 91125, USA}
\date{Received; accepted; published online}

\abstract{9C J1503+4528 is a very young CSS radio galaxy, with an age
  of order $10^{4}$ years.  This source is 
  an ideal laboratory for the study of the intrinsic host galaxy/IGM
  properties, radio source interactions, evidence for young stellar
  populations and the radio source triggering mechanism. Here we present
  the results of a spectroscopic analysis of this source, considering
  each of these aspects of radio source physics.
}

\correspondence{k.inskip@shef.ac.uk}

\maketitle

\section{Introduction}
It is now widely accepted that Compact Steep Spectrum (CSS) and
Gigahertz Peaked Spectrum (GPS) radio sources are in the earliest
stages of their evolution, being the progenitors of the larger FRII
source population (e.g. Fanti 1995, Readhead 1996). Due to synchrotron
self-absorption, the radio spectra of sources with ages of $\sim 1000$
years will peak at frequencies of between $\sim 1$GHz and $\sim 10$GHz
(O'Dea 1998; Snellen et al 2000).  Spectral aging arguments
(e.g. Murgia et al 1999) suggest ages as young as $< 10^4$ years for
some sources, whilst VLBI-based dynamical ages of GPS sources are
$\sim 10^3$ years (e.g. Owsianik, Conway \& Polatidis 1998; Taylor et
al 2000).

However, our understanding of this early phase of radio galaxy evolution 
is hampered by the difficulty of compiling uniform samples of
radio sources in the first few millennia of their growth, and the
impossibility of pinpointing galaxies immediately
prior to this stage, whilst the processes triggering the radio jets are 
still underway. 

The 9C survey (Waldram et al 2003) was carried out at 15 GHz with the
Ryle Telescope to identify foreground sources for the Very Small Array
(VSA) microwave background experiment.  Over the past several years,
extensive followup of the radio properties of this population has been
carried out (Bolton et al 2003, 2004; Bolton et al in prep), both to
better understand the effects on microwave background experiments and
to investigate the sources themselves.

This high survey frequency has been proved to select significant numbers 
of CSS and GPS sources, as well as sources peaking in the $\sim 10$ GHz 
range.  We believe this is due to two underlying reasons.  First, any 
survey will preferentially select sources which are brightest near the 
survey frequency; hence, we preferentially select sources peaking at high 
frequency. Second, and more subtly, the youngest CSS sources will have 
little spectral aging, and thus will be brighter at high frequency than 
their older counterparts.

\begin{figure*}
\centering
\resizebox{12.1cm}{!}
{\includegraphics[angle=90]{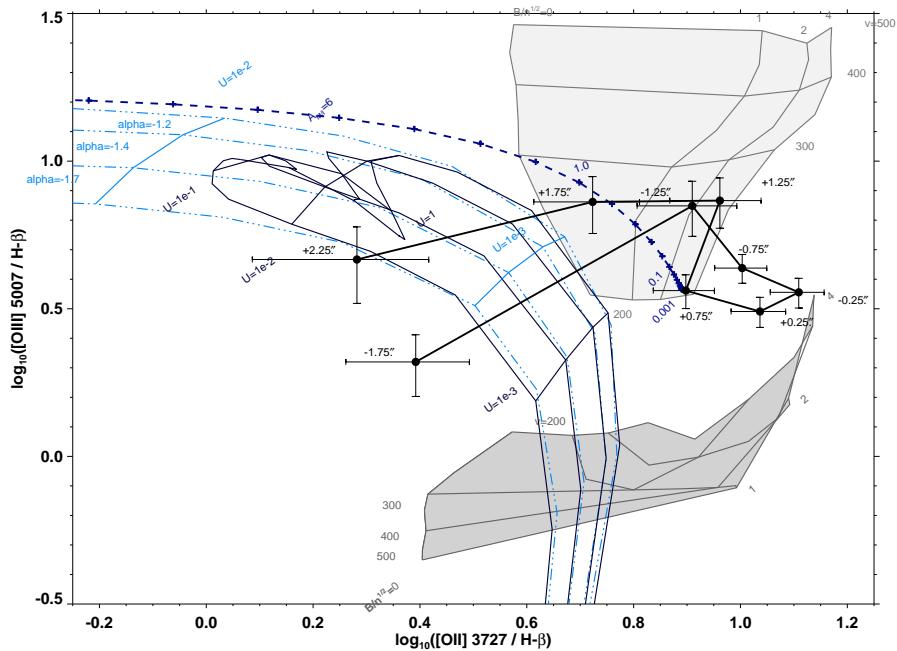}}
\caption{Ionization mechanism diagnostic plot for 9C J1503+4528, using
  the line ratios [O\textsc{ii}]/H$\beta$ and [O\textsc{iii}]/H$\beta$. Data
  points represent spectra extracted in 0.5$^{\prime\prime}$ steps,
  and are labelled in terms of their distance from the continuum
  centroid.  Negative positions are towards the NW, and positive positions
  towards the SE.  The model tracks 
  include: (1) shock 
  ionization (Dopita \& Sutherland 1996), with and
  without a precursor photoionzation region (light \& dark shading
  respectively), (2) Simple 
  AGN photoionization tracks (light grey, triple-dot-dashed lines), including
  the effects of 
  dust (dark grey, solid lines, Groves, Dopita \& Sutherland
  2004a,b), and (3) the mixed-medium matter-bounded vs. ionization-bounded photoionization model (dashed
  track) of Binette,
  Wilson \& Storchi-Bergmann (1996).}
\label{ionize1}
\end{figure*}

Optical followup of the CSS and GPS sources is now underway.  Here we
present the initial results of deep optical spectroscopy of one such
source. 9C J1503+4528 is a classical double CSS source at $z = 0.521$,
with an angular size of $0.5^{\prime\prime}$, corresponding to a
projected linear size of $\approx 3$ kpc.  

Values for the cosmological parameters of $\Omega_0 = 0.27$,
$\Omega_{\Lambda} = 0.73$ and $H_0 = 65\, \rm{km\,s}^{-1}$ are assumed
throughout.

\section{Features of interest}
Spectra of 9C J1503+4528 were obtained using Keck LRIS (Oke et al
1995) with a 1$^{\prime\prime}$ slit aligned both parallel and
perpendicular to the radio source axis, and integration times of 1800s
for each position angle.  Powerful line emission is observed, with the
strength of the \oo3727\AA\, line cf [O\textsc{iii}](4959+5007)\AA\, indicating a
fairly low ionization state for the majority of the gas. The line
emission is fairly broad in the central regions of both spectra; a
narrow component is also observed in the outer regions of the galaxy
in the spectrum aligned parallel with the radio source.  The extent of
the narrow component ($\approx 5^{\prime\prime}$) is roughly ten times
larger than that of the radio source itself, but not large compared to
the extended emission line regions (EELRs) observed around other radio
sources (e.g. Inskip et al 2002a, Best, R\"{o}ttgering \& Longair
2000). We therefore infer that the ISM/IGM is being ionized by the by
the AGN, out to a limiting distance of $c$ times the age of the AGN
itself.  Taking a typical growth rate of a compact radio source to be $
\sim 0.1c$ (e.g. Owsianik, Conway \& Polatidis 1998; Taylor et al
2000), the factor of $\approx 10$ difference between radio source size
and EELR size implies that both the AGN and the radio source were
triggered at effectively the same point in time, with the source age
being $\sim 10^4$ years.

Determining the systemic redshift of 9C J1503+4528 was complicated by
the inherent structure in the emission lines: the redshift implied by
the peak of the line emission on the continuum centroid differs
substantially from that of the narrow extended component.  However,
the various absorption features in the spectra (Ca H \& K, high-order
Balmer lines) all suggest a redshift identical to that determined from
the extended narrow component, giving a value of $z = 0.521$.

The presence of the Balmer absorption features, together with blue
continuum emission ({\it see Inskip et al in prep. for further
details}), indicate the presence of a young stellar population (YSP).
Whilst the line emission is more pronounced in the parallel spectrum,
absorption is more significant in the perpendicular spectrum,
suggesting that any YSP is distributed throughout the galaxy rather
than being associated with any radio source shocks.

\section{Ionization state}

Ionization state diagnostic diagrams (Baldwin, Phillips \& Terlevich
1981) are particularly useful in investigating emission line gas
(e.g. Best, R\"{o}ttgering \& Longair 2000, Inskip et al 2002a, Moy \&
Rocca-Volmerange 2002).  To investigate the changing ionization state
of the emission line gas with position, one-dimensional spectra were
extracted in $0.5^{\prime\prime}$ steps across the slit from the
parallel spectrum, and the emission lines fitted by gaussians.  The
measured line ratios were compared with the predictions of various
models.  Figure \ref{ionize1} displays the \oo3727\AA/H$\beta$
vs. [O\textsc{iii}]5007\AA/H$\beta$ diagnostic diagram (see figure caption for
details).

The changing ionization state relative to distance from the AGN can be
immediately appreciated.  The central regions of the host galaxy lie
close to the predictions of the shock ionization models (and also the
far end of the mixed medium ionization track).  Just outside the radio
source, the shock plus precursor photoionization models provide a good
explanation.  Beyond 1-1.5$^{\prime\prime}$ (i.e. the extended narrow
emission region), the data are well described by the predictions of
simple AGN photoionization.  The higher values of the ionization
parameter are most plausible, so dusty models are clearly more
appropriate than photoionization in the absence of dust.  For the gas
which is (apparently) AGN photoionized, the ionization state of the
material at larger radii suggests either a lower spectral index
($\alpha \sim -2.0$ rather than $\alpha \sim -1.2$), higher gas
density or lower metallicity.  Although the mixed medium model also
seems to provide a good representation of the changing emission line
ratios with position, its predictions do not coincide with the
behaviour of the [O\textsc{iii}]4363\AA/[O\textsc{iii}]5007\AA\, line ratio, allowing us to
exclude this type of model from further consideration.

The results for the central 1-2$^{\prime\prime}$ should not be greatly
different for the parallel and perpendicular spectra, as both sample
essentially the same region of the host galaxy.  However, the impact
of AGN and shock ionization will be greatly lessened at larger
distances, for material which lies off the radio source axis and well
outside the ionization cone of the AGN. This is indeed observed:
similar line ratios are observed within the central
2$^{\prime\prime}$, but the line emission at a greater distance from
the AGN is much weaker for the perpendicular spectrum, and the
emission line ratios for the ionized component are consistent with a
weaker flux of ionizing photons.

The picture provided by these results is as follows.  The UV continuum
emission from the AGN ionizes the gas at all locations within the
ionization cone, decreasing in intensity at larger distances. Much
weaker line emission is observed off-axis. The innermost regions of
the galaxy, of order the same size as the radio source, have an
additional contribution from shocks.  The UV radiation field produced
by these shocks adds another ionization component to the emission from
gas at intermediate distances.

\section{Kinematics}
We have also investigated the gas kinematics by analysing the profiles
of the \oo 3727\AA\, and  [O\textsc{iii}](4959+5007\AA) emission lines.
Two--dimensional regions around each of these lines were
extracted. From these, a sequence of one dimensional spectra were
extracted along the slit direction, stepped every
0.5$^{\prime\prime}$. The extracted spectra were then analysed using
the following procedure:

After continuum subtraction, the data were fitted by successive
Gaussian components, which were only accepted if their FWHM was larger
than the instrumental resolution and the S/N ratio was greater than
approximately five (with the exception of the outermost extracted
spectra, for which a fainter, single gaussian component could be
detected at a lower S/N.)  The best-fit combination of Gaussians (that
with the lowest reduced $\chi^{2}$) determined the maximum number of
velocity components which could realistically be fitted to the data.
(For the [O\textsc{iii}]\, emission line, the relative positions and strengths of
the two components are held fixed)

The integrated emission line flux, the velocity relative to that at
the centre of the galaxy, and the emission line FWHM were determined
for each Gaussian fit. To calculate the emission line FWHM,
it was deconvolved by subtracting in quadrature the
instrumental FWHM, as determined from unblended sky lines. 
Errors were calculated for these three parameters, and allow for the
fact that a range of possible fits are equally acceptable at low S/N.
This approach allows us to search for high velocity components in the
emission line gas, or other structures incompatible with a fit to a
single velocity component.  Given the observed emission line profile
of this source, such a step was a necessity, and broader emission line
components are clearly present in the central parts of the extended
emission line region.

\begin{figure}
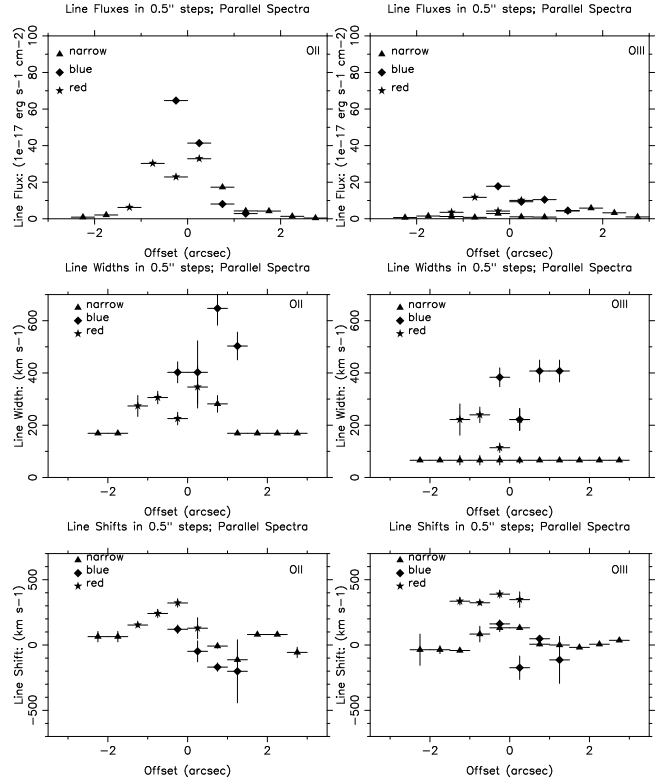

\resizebox{\hsize}{!}
{\includegraphics[angle=-90]{OIIflux.ps}
\includegraphics[angle=-90]{OIIIflux.ps}}\\
\resizebox{\hsize}{!}
{\includegraphics[angle=-90]{OIIwidth.ps}
\includegraphics[angle=-90]{OIIIwidth.ps}}\\
\resizebox{\hsize}{!}
{\includegraphics[angle=-90]{OIIshift.ps}
\includegraphics[angle=-90]{OIIIshift.ps}}
\caption{Gas kinematics for 9C J1503+4528.  The changing line fluxes
  (top), widths (centre) and offsets (bottom) with position are
  displayed for the \oo (left) and [O\textsc{iii}] (right) emission lines,
  extracted and fitted in 0.5\arcsec steps from the parallel
  spectrum. Up to three emission line components are modelled at each
  position.  Negative offsets are towards the NW, and positive offsets
  towards the SE.}
\label{K1}
\end{figure}

Our results for the kinematics of the parallel spectrum are presented
in Fig.~\ref{K1}, which displays flux, width and velocity offset for
fitted line components at various positions across the spectrum, for
both \oo and [O\textsc{iii}].

An  unresolved emission component is present in the outer
regions of the EELR, in both \oo\, and [O\textsc{iii}], at wavelengths which are
generally consistent with the emitting material lying at rest relative 
to the host galaxy.  This narrow component can be traced across
the galaxy in [O\textsc{iii}].  The \oo emission line is of substantially higher
luminosity, particularly in the central regions of the galaxy.
Because of this, the narrow component of \oo\, can only be observed in
the outer regions, and is lost in the centre in favour of one or more
broader components. The relative strengths of the narrow components of
\oo and [O\textsc{iii}] are consistent with simple AGN photoionization, as
demonstrated by our diagnostic diagrams in the previous section.

The broader emission line components extend to a distance of
approximately 1-1.5$^{\prime\prime}$ from the continuum centroid. We
note that the \oo emission is substantially stronger than the [O\textsc{iii}]
emission; such a low ionization state in the emitting material
suggests shocks associated with the radio source, in which case the
emitting material would display larger line widths.  Usually a single
broad component is sufficient, except in the central regions where two
broad components are preferred.  The broad components on the SE side
of the galaxy usually lie at shorter wavelengths, with a velocity
shift of a few hundred kms$^{-1}$, whilst the emission from the
opposite side of the galaxy is shifted to longer wavelengths by a
similar amount.  Of these two distinct components, the blue-shifted
emission line component is generally more luminous, and has a larger
line width (400-600 km s$^{-1}$, cf 200-400 km s$^{-1}$ for the
red-shifted emission).  It is hard to say whether these results
suggest outflows or rotation of the gas.

The emission lines in the perpendicular spectrum do not show an
extended narrow component.  This is strong evidence that the source of
the ionizing photons for the extended narrow component is the
ionization cone of the AGN, which does not intersect with the
perpendicular spectrum except in its centre.  In the central regions
of the perpendicular spectrum, broad emission components are observed.
These are very similar to those observed in the parallel spectrum,
with the exception that there is no clear tendency for the blue- or
red-shifted emission to preferentially lie on one side of the galaxy
or the other.  This suggests that outflows along the radio axis are a
plausible option for explaining the gas kinematics, as an alternative
to simple rotation.

\section{Conclusions}
The spectra of the very young radio source 9C J1503+4528 have produced a
number of interesting results. 
\vspace{-0.2cm}
\begin{enumerate} 
\item[$\bullet$]The relative sizes of EELR and radio source (the
  former being $\approx 10$ times larger), together with typical
  compact radio source expansion velocities of $\sim0.1c$, suggest
  that both the luminous quasar nucleus responsible for photoionizing
  the outer regions of the EELR, and the radio source itself, were
  triggered more-or-less simultaneously.
\item[$\bullet$]The EELR ionization state is consistent with shocks
  being important on scales comparable to the radio source, with AGN
  photoionization within the ionization cones being the dominant
  mechanism further from the AGN.
\item[$\bullet$]The gas kinematics confirm that weak shocks could be
  important close to the radio source, and that the outer regions of
  the EELR are relatively undisturbed.  The kinematics suggest that
  motion along the radio axis may be taking place, which may indicate
  the presence of outflows.
\item[$\bullet$]The blue continuum emission and  balmer absorption
  features suggest that recent star formation has occurred, but do not
  imply that it is directly related to the radio source.
\end{enumerate}

We have found 9C J1503+4528 to be an excellent example for the
detailed study of the properties of young radio sources, and we expect
that studies of comparably selected sources will continue to shed
light on the processes surrounding radio source triggering.
The observed changes in gas kinematics and
dominant ionization mechanism in different regions of the EELR are not
unexpected, given the expected location of radio source shocks, but do
provide two key results: the clumpy ISM/IGM responsible for the EELRs
observed around powerful radio sources is indeed in place prior to the
expansion of the radio source, and is relatively undisturbed in terms
of its kinematic properties.

If the triggering of radio source activity in this galaxy
was due to a merger or interaction, the properties of any young stellar
population formed as part of that process can shed light on the
exact mechanisms and time scales involved.  Continuum modelling of
these spectra (Inskip et al in prep) suggests that the YSP is likely
to be of a fairly low total stellar mass, and relatively young
(although still substantially older than the radio source). These
results, together with the relatively undisturbed
appearance of the host galaxy and the quiescent kinematics in the
outer regions of the EELR suggest that any merger/interaction
responsible for the eventual triggering of the radio source activity
was either fairly minor in nature, or occurred a considerable time
previous to the epoch of the observations. 

\acknowledgements{The data presented
herein were obtained at the W. M. Keck Observatory, which is operated
as a scientific partnership among the California Institute of
Technology, the University of California, and the National Aeronautics
and Space Administration. The Observatory was made possible by the
generous financial support of the W. M. Keck Foundation. 
KJI acknowledges a PPARC research fellowship, DL a PPARC PhD
studentship and GC support from PPARC observational rolling grant
PPA/G/O/2003/00123.
It is a pleasure to express our gratitude and respect to the
indigenous people of Hawai`i, from whose sacred mountain Mauna Kea our
observations were made. 

 }

\end{document}